\RequirePackage{fix-cm}
\documentclass{svjour3}                     % onecolumn (standard format)
\smartqed  % flush right qed marks, e.g. at end of proof
\usepackage{graphicx}
\newcommand {\be} {\begin{equation}}
\newcommand {\ee} {\end{equation}}
\newcommand {\Be}{\begin{eqnarray*}}
\newcommand {\Ee} {\end{eqnarray*}}
\newcommand {\bey} {\begin{eqnarray}}
\newcommand {\eey} {\end{eqnarray}}
\newcommand{\bit}{\begin{itemize}}      
\newcommand{\eit}{\end{itemize}}
\newcommand{\bfl}{\begin{flusleft}}
\newcommand{\efl}{\end{flusleft}}
\newcommand{\bfr}{\begin{flushright}}
\newcommand{\ec}{\end{center}}
\newcommand{\ben}{\begin{enumerate}}    
\newcommand{\een}{\end{enumerate}}

\newcommand{\comment}[1]{}

\begin{document}

\title{Stability of the splay state in networks of pulse-coupled neurons}

\author{Simona Olmi \and Antonio Politi \and Alessandro Torcini}

\institute{%
Simona Olmi, Antonio Politi, Alessandro Torcini \at
CNR - Consiglio Nazionale delle Ricerche,
Istituto dei Sistemi Complessi, via Madonna del Piano 10, I-50019 Sesto
Fiorentino, Italy\\
Centro Interdipartimentale per lo Studio delle Dinamiche Complesse, via Sansone, 1 
- I-50019 Sesto Fiorentino, Italy\\
\and Antonio Politi \at
Institute for Complex Systems and Mathematical Biology, King's College,
University of Aberdeen, Aberdeen AB24 3UE, United Kingdom
\and Simona Olmi, Alessandro Torcini \at
INFN Sez. Firenze, via Sansone, 1 - I-50019 Sesto Fiorentino, Italy\\
\email{simona.olmi@fi.isc.cnr.it, a.politi@abdn.ac.uk, alessandro.torcini@cnr.it}
}%

\maketitle

%%%%%%%%%%%%%%%%%%%%%%%%%%%%%%%%%%%%%%%%%%%%%%%%%%%%%%%%%%%%%%%%%%%%%%%%%%%

\begin{abstract}
We analytically investigate the stability of {\it splay states} in networks of
$N$ pulse-coupled phase-like models of neurons. By developing a perturbative
technique, we find that, in the limit of large $N$, the Floquet spectrum scales
as $1/N^2$ for generic discontinuous velocity fields. Moreover, the stability of
the so-called short-wavelength component is determined by the sign of the jump
at the discontinuity. Altogether, the form of the spectrum depends on the pulse
shape but is independent of the velocity field.
\end{abstract}

%%%%%%%%%%%%%%%%%%%%%%%%%%%%%%%%%%%%%%%%%%%%%%%%%%%%%%%%%%%%%%%%%%%%%%%%%%%%%

\section{Introduction}
The first objective of a (neural) network theory is the identification of the
asymptotic regimes. The last-decades activity have led to the discovery of
fully- and partially-synchronized states, clusters and splay or asynchronous
states in pulse-coupled networks \cite{mir_synch,abbott,vvres,hansel}. It has
also been made clear that ingredients such as disorder (diversity of the
neurons and structure of the connections) are very important in determining
the asymptotic behaviour, as well as the possible presence of delayed
interactions and plasticity \cite{gerstner,dayan}.
However, even if one restricts the analysis to identical, globally-coupled
oscillators, there are very few theoretical results: they mostly concern
fully synchronized regime or specific types of neurons (e.g. the leaky
integrate-and-fire model)~\cite{hansel,brunel2000,ermentrout}.

In this paper, we develop a perturbative analysis for the stability of
{\it splay states} (also known as antiphase states~\cite{splay},
``ponies on a merry-go-round"~\cite{krupa}, or rotating waves~\cite{ashw90}) 
in ensembles of $N$ identical fully pulse-coupled 
neurons. In a splay state, all the neurons follow the same periodic dynamics
except for a time shift that is evenly distributed. Splay states have been
identified in experimental measurements performed on electronic
circuits~\cite{ashw90} and on multimode lasers
\cite{wiese90}. Theoretical studies have been devoted to splay states
in fully coupled Ginzburg-Landau equations \cite{hakim92}, Josephson arrays
\cite{nicols}, laser models \cite{rappel94}, traffic models \cite{seidel}, 
and pulse-coupled neuronal networks \cite{abbott}. 
In the latter context, splay states have been mainly investigated in
leaky-integrate-and-fire (LIF) neurons
\cite{abbott,vvres,bressloff99,zillmer2}, but some studies have been also
devoted to the $\theta$-neurons~\cite{dipoppa} and to more realistic neuronal
models~\cite{brunel_hansel06}.
Finally, splay states are important  in that they provide the simplest instance
of asynchronous behaviour and can be thereby used as a testing ground for the
stability of a more general class of dynamical regimes.

Our model neurons are characterized by a membrane potential $u$ that is
continuously driven by the velocity field $F(u)$, from the resetting value $u=0$
towards the threshold $u=1$ (see the next section for a more precise
definition). As threshold and resetting value can be identified
with one another and thereby $u$ interpreted as a phase, it will be customary to
refer to the case $F(1)\ne F(0)$ as to that of a discontinuous velocity field.
Additionally, we assume that the post-synaptic potential (PSP) has a stereotyped
shape, the so-called $\alpha$-pulse, that is characterized by an identical rise
and decay time $1/\alpha$ \cite{abbott}. As already discussed in \cite{zillmer2},
the Floquet spectrum is composed of two components: (i) long wavelengths (LWs),
which can be studied in terms of a suitable functional equation for the
probability distribution of the membrane potential $u$ \cite{abbott}; (ii)
short-wavelengths (SWs), which typically correspond to marginally stable
directions in the thermodynamic limit ($N\to\infty$). By developing an approach
that is valid for arbitrary coupling strength and is perturbative in the inverse
system-size $1/N$, we prove that the SW component of the Floquet spectrum scales
as $1/N^2$ and is proportional to $F(1)-F(0)$, i.e. it is present
only if the velocity field is discontinuous. We are also able to determine the
spectral shape and find it to be {\it universal}, i.e. independent of the
details of the velocity field. 

More precisely, we first build the corresponding event-driven map, by expanding 
it in powers of $1/N$ (a posteriori, we have verified that it is necessary to
reach the fourth order). Afterwards, the expression of the splay state is
determined: this task corresponds to finding a fixed point of the event-driven
map in a suitably moving reference frame - analogously to what previously done
in specific contexts \cite{zillmer2,calamai,olmi}. In practice this task is
carried out by first taking the continuum limit for the various orders
and obtaining suitable differential equations, whose solution allows proving
that all finite-size corrections for both the period $T$ and the membrane
potential vanish up to the third order. Next, the stability analysis is carried out
to determine the leading term to the Floquet spectrum. This task involves the
introduction of a suitable Ansatz to decompose each eigenvector into the linear
superposition of a slow and a rapidly oscillating component. The following
continuum limit shows that the two components satisfy an ordinary and a
differential equation, respectively.

Altogether, the proof of our main result requires determining all terms up to
the third order in the $1/N$ expansion of the splay state solution, while some third order terms
are not necessary for the tangent space analysis. Going beyond discontinuous
fields would require extending our analysis to account for higher order terms
and this might not even be sufficient to characterize analytic velocity fields.
In fact, previous numerical simulations \cite{calamai} suggest that the Floquet
exponents scale with higher powers of $1/N$ that depend on which derivatives of
$F(u)$ are eventually discontinuous. Moreover, it is worth
recalling that in the case of a strictly sinusoidal field, a theorem proved by
Watanabe and Strogatz~\cite{watanabe} implies that $N-3$ Floquet exponents
($N-2$ for a splay solution) vanish exactly for any value of $N$.

In the small coupling limit, one can combine our results with those of Abbott
and van Vreeswijk \cite{abbott} (that are valid only in that regime) for the LW
spectral component and conclude that the splay state is stable whenever
$F(0) > F(1)$ and the pulses are sufficiently broad, for excitatory coupling,
while it is always unstable for inhibitory coupling and any finite pulse-width.
This scenario is partially reminiscent of the stability of synchronous and
clustered regimes that is determined by the sign of the first derivative
$dF/du$ of the
velocity-field averaged on the interval $[0,1]$ (this latter problem has
been investigated in excitatory pulse-coupled integrate-and-fire oscillators
subject to $\delta$-pulses \cite{mir_synch,mauroy}).

Section II is devoted to the introduction of the model and to a brief
presentation of the main results, including an expression for the leading
correction to the period for the LIF model, to provide evidence that they
are typically of 4th order.
A general perturbative expression for the map is derived in Sec. III, while
Sec. IV is devoted to deriving the splay-state solution up to the third order in
$1/N$.
The main result of the paper is discussed in Sect. V, where the Floquet spectra
are finally obtained. Sect. VI contains some general remarks and 
a discussion of the open problems. The technical details of some
lengthy calculations have been confined in the appendices: Appendix A is
devoted to the derivation of the splay state solution; Appendix B contains the
derivation of the leading term (of order four) of the period $T$ for the LIF
model; Appendix C is concerned with the linear stability analysis.

%%%%%%%%%%%%%%%%%%%%%%%%%%%%%%%%%%%%%%%%%%%%%%%%%%%%%%%%%%%%%%%%%%%%%%%%%%%%%

\section{Model and main results}\label{two}

We consider a network of $N$ identical neurons (rotators) coupled via a
mean-field term. The dynamics of the $i$-th neuron writes as
\begin{equation}
\label{eq:x1}
  \dot{u}_i(t)= F(u_{i})+gE(t) \equiv {\mathcal F}_i(t) \, \qquad i=1,\dots,N
\qquad ,
\end{equation}
where $u_i(t)$ represents the membrane potential, $E(t)$ is the
forcing field, and $g$ is the coupling constant. When the membrane potential
reaches the threshold value $u_i(t)=1$, a spike is sent to all neurons
(see below for the relationship between the single spikes and the global
forcing field $E(t)$) and it is reset to $u_i(t)=0$. The resetting procedure
is an approximate way to describe the discharge mechanism operating in real
neurons. The function $F$ represents a velocity field for the isolated
neuron and it is assumed to be everywhere positive (thus
ensuring that the neurons repetitively fire, since they are supra-threshold),
while ${\cal F}_i$ is the velocity field seen by the neuron $i$ in the 
presence of a coupling with other neurons. While we consider both
excitatory ($g>0$) and inhibitory networks ($g <0$), it is easy to show 
that ${\cal F}$ remains always positive to ensure the existence of
splay states. For the simple choice
\begin{equation}
\label{eq:LIF}
F(u)=a-u
\qquad ,
\end{equation}
the model reduces to the well known case of LIF neurons.

The field $E$ is the linear superposition of the pulses emitted in the past
when the membrane potential of each single neuron has reached the threshold
value. By following Ref.~\cite{abbott}, we assume that the shape of a pulse
emitted at time $t=0$ is given by
$E_s(t)= \frac{\alpha^2 t}{N} {\rm e}^{-\alpha t}\,$, where $1/\alpha$
is the pulse--width. This is equivalent to saying that the total field evolves
according to the equation
\begin{equation}
\label{eq:E}
  \ddot E(t) +2\alpha\dot E(t)+\alpha^2 E(t)= 
  \frac{\alpha^2}{N}\sum_{n|t_n<t} \delta(t-t_n) \ .
\end{equation}
where the sum in the r.h.s. represents the source term due to the spikes
emitted at times $t_n<t$.

It is convenient to transform the continuous-time model into a discrete-time
mapping. We do so by integrating the equations of motion from
time $t_n$ to time $t_{n+1}$ (where $t_n$ is the time immediately after the
$n$-th pulse has been emitted). The resulting map for the field variables reads,
\begin{eqnarray}
\label{eq:map}
  E_{n+1}&=& [E_n+ \tau_n P_n] {\rm e}^{-\alpha \tau_n} \quad ,
\\ \nonumber
  P_{n+1}&=& P_n {\rm e}^{-\alpha \tau_n}+\frac{\alpha^2}{N}\ ,
\end{eqnarray}
where $\tau_n= t_{n+1}-t_n$ is the interspike time interval and, 
for the sake of simplicity, we have introduced the new variable
$ P := \alpha E+\dot E $.

In this paper we focus on a specific solution of the networks dynamics,
namely on \textit{splay states}, which are asynchronous states, where all neurons 
fire periodically with period $T$ and two successive spike emissions occur
at regular intervals $\tau_n \equiv T/N $. The first result of this paper
is that under the assumption that the velocity
field $F(u)$ is differentiable at least four times, the dependence of the
period $T$ onto the size $N$ is of order $o(1/N^3)$. In the specific case of
LIF neurons, we show in Appendix B that the leading correction $\delta T$ to
the infinite size result is indeed of order $O(1/N^4)$ and, more precisely,
\begin{equation}
\label{eq:T4LIF}
\delta T= \frac{K(\alpha)-6  }{720}
\frac{\left[a(1 - {\rm e}^{-T}) -1\right]}
    { ge^{-T} + a\left( T  + 1 - {\rm e}^{-T}\right) -1}
    \ \frac{T^{5}}{N^4}
\qquad ,
\end{equation}
where $K(\alpha)$ encodes the information on the pulse dynamics (see
Eq.~(\ref{eq:Kdef})).
We did not dare to estimate the quartic contribution for
generic velocity fields, not only because the algebra would be utterly
complicated, but also since our main motivation is to determine the leading
contributions in the stability analysis, and it turns out that it is sufficient
to determine the splay state up to the third order.

The study of the stability requires determining the Floquet spectrum, i.e.
the complex eigenvalues of a given periodic orbit of period $T$. With reference
to a system of size $N$, the Floquet multipliers can be written as
\begin{equation}
 \mu_k = {\rm e}^{i \phi_k} {\rm e}^{(\lambda_k + i \omega_k)\tau_n}
\qquad , \quad k=1, \dots, 3N \qquad ,
\label{floquet}
\end{equation}
where $\phi_k$ represents the 0th order phase (that is responsible for the
high frequency oscillations of the corresponding eigenvector - see Sec. V),
while $\lambda_k$ and $\omega_k$ are the real and imaginary parts of the 
Floquet exponent, respectively.
In the following we prove that the leading term of the SW component
(i.e. for $\phi_k$ away from  zero), is
\begin{equation}
\label{eq:40}
\lambda_k = \frac{g\alpha^2}{12}
\frac{F(1) - F(0)}{\mathcal{F}(1)\mathcal{F}(0)}
\left(\frac{6}{1-\cos\phi_k}-1\right) \frac{1}{N^2}
\qquad .
\end{equation}
For discontinuous velocity fields, the real parts of the spectrum scale as
$1/N^2$, while the imaginary parts are of even higher order.

For continuous fields, it has been numerically observed  that 
the scaling of the spectrum is at least $O(1/N^4)$ \cite{calamai}.
In other words the shape of the spectrum is universal, apart from
a multiplicative factor that vanishes if and only if $F(1)=F(0)$,
i.e. for true phase rotators where $u=0$ coincides with $u=1$.
The stability of the splay state can be inferred by 
the sign of $F(1) - F(0)$: in the case of excitatory (resp. inhibitory)
coupling, the state is stable whenever $F(0) > F(1)$ (resp. $F(0) < F(1)$). 
In the limit $\phi_k \to 0$ the expression reported in parenthesis in
Eq.~(\ref{eq:40}) diverges, indicating that the perturbative analysis breaks
down. This limit corresponds to the LW component, where our approach can be
complemented by that of Abbott and van Vreeswijk~\cite{abbott}, which reveals
that the corresponding Floquet exponents do not depend on the system size.
For sufficiently small couplings ($|g| << 1$), they also found a condition
similar to the one reported above, namely that, irrespectively of the 
sign of the coupling, the splay state is stable whenever $F(0)>F(1)$ for
sufficiently broad pulses. In fact, above a critical $\alpha$-value (i.e. below
a given pulsewidth), the splay state looses stability due to a supercritical
Hopf bifurcation, which leads to the emergence of a more complex collective
regime, termed {\it partial synchronization} \cite{vvres,mohanty}.
By combining the conditions for the SW and the LW spectrum, one can predict
the overall stability of the splay state. In particular, the state is stable for
excitatory coupling if $F(0) > F(1)$ (and $\alpha$ sufficiently small), while
it is always unstable for finite networks,for inhibitory coupling, since the 
SW and LW stability conditions are opposite to one another. This last result is
consistent with the findings reported by van Vreeswijk for inhibitory coupling
and finite pulse width~\cite{vvres}.

\section{Event driven map}

By following Ref.~\cite{zillmer,calamai}, it is convenient to pass from a
continuous to a discrete time evolution rule, by introducing the event-driven
map which connects the network configuration at subsequent spike emissions
occurring at time $t_n$ and $t_{n+1}$. The membrane-potential value
$u_i(t^-_{n+1})$ just before the emission of the $(n+1)$-th spike can be obtained
by formally integrating Eq.~(\ref{eq:x1}),
\begin{equation}
\label{eq:plus} 
u_i(t^-_{n+1}) - u_{n,i}(t_n) =
\int_{t_{n}}^{t_{n+1}^-} dt F(u_{i}(t)) +
g\int_{t_{n}}^{t_{n+1}^-} dt \big[E_n+ P_n(t- t_n)\big] {\rm e}^{-\alpha t}
\equiv {\mathcal A}_1 + {\mathcal A}_2 \, ,
\end{equation}
where the minus superscript means that the map construction has not yet been
completed. This task is accomplished by ordering the membrane potentials 
from the largest ($j=1$) to the smallest value ($j=N$) value and by passing to
a comoving frame that advances with the firing neuron, 
i.e. by shifting the neuron index by one unit,
\begin{equation}
\label{eq:int} 
u_{n+1,j-1} = u_{j}(t^-_{n+1}) \, ,
\end{equation}
where the first subscript indicates that the variable is determined at time
$t_{n+1}$.  This change of reference frame allows treating the splay state as
a fixed point  of the event driven map.

The first integral appearing on the rhs of Eq.~(\ref{eq:plus}) is now solved
perturbatively by introducing a polynomial expansion of $u_i(t)$ around $t=t_n$,
which, up to third order, reads as
\begin{equation}
\label{eq:expa1}
u_{j}(t)= u_{n,j} + \dot {u}_{n,j}\delta t
 + \frac{1}{2}\ddot{u}_{n,j}\delta t^2
+ \frac{1}{6}{u}^{\hspace{-0.25cm}\cdots}_{n,j} \delta t^3 + O\left(\delta
t^{4}\right) \quad,
\end{equation}
where $\delta t = t - t_n$.
Explicit expressions for the time derivates of
$u_j$ can be obtained from Eq.~(\ref{eq:x1}) and its time derivatives,
\begin{eqnarray}
\nonumber
%%\label{eq:ddots}
\ddot{u}_{n,j}=F'\left( {u}_{n,j} \right) \dot{u}_{n,j} + g\dot{E}_n 
\quad ,
\\ \nonumber
{u}^{\hspace{-0.25cm}\cdots}_{n,j} = F''\left( {u}_{n,j}\right)\dot{u}_{n,j}^2 
+ F'\left( {u}_{n,j}\right) \ddot{u}_{n,j} + g\ddot{E}_n \quad ,
\end{eqnarray}
where one can further eliminate $\ddot{E}_n$ with the help of Eq.~(\ref{eq:E}).

By inserting the expansion (\ref{eq:expa1}) into the expression of
${\mathcal A}_1$, expanding the function $F(u)$, and performing the
trivial integrations, one obtains
\begin{eqnarray} 
\label{int1}
\hspace{-0.5cm}
&&{\mathcal A}_1 = F_{n,j}\tau_n + F'_{n,j} {\mathcal F}_{n,j} \frac{\tau_n^2}{2} +
\left\{ 
\Big[
F''_{n,j} {\mathcal F}_{n,j} +  F_{n,j}'^2
\Big]
{\mathcal F}_{n,j} + g  \dot{E}_n F'_{n,j} 
\right\}
\frac{\tau_n^{3}}{6} +
\left\{ 
F'''_{n,j} {\mathcal F}^3_{n,j}
\right.
\\  
\nonumber
&& \left.
%\hspace{-0.5cm}
+
4 F'_{n,j} F''_{n,j} {\mathcal F}^2_{n,j} +
F_{n,j}'^{3} {\mathcal F}_{n,j} + 
g \left[
\Big(
3 F''_{n,j} {\mathcal F}_{n,j} + F_{n,j}'^2 -  \alpha F'_{n,j}
\Big)
\dot{E}_n - \alpha   F'_{n,j}  P_n
\right]   
\right\}
\frac{\tau_n^4}{24} + O(\tau_n^5) \, ,
\nonumber
\end{eqnarray}
where $\tau_n = t_{n+1}-t_n$ and we have introduced the short-hand notation
$F_{n,j}$ for $F(u_{n,j})$ (and analogously for $\mathcal F$).

The explicit expression of $\mathcal A_2$ reads
\begin{eqnarray}
\label{int2}
{\mathcal A_2} &=&
\frac{g}{\alpha}E_n(1-{\rm e}^{-\alpha\tau_n}) -
\frac{g}{\alpha}P_n\tau_n {\rm e}^{-\alpha\tau_n} 
+ \frac{g}{\alpha^{2}}P_n(1-{\rm e}^{-\alpha\tau_n})
\\
\nonumber
&=& g E_n \tau_n + g\dot{E}_n \frac{\tau_n^2}{2} - 
g \alpha \Big( \dot{E}_n + P_n \Big)\frac{\tau_n^3}{6} +
g \alpha^2 \Big( \dot{E}_n + 2P_n \Big)\frac{\tau_n^4}{24} + O(\tau_n^5) \, .
\end{eqnarray}
Now, by assembling Eqs.~(\ref{eq:plus},\ref{eq:int},\ref{int1},\ref{int2}), we
obtain the final expression for the evolution rule of the membrane potential,
\begin{eqnarray}
\label{eq:tot}
&& u_{n+1,j-1} = u_{n,j} + \mathcal{F}_{n,j}\tau_n + 
\left[
g{\dot E}_n + F'_{n,j} {\mathcal F}_{n,j}
\right]
\frac{\tau_n^2}{2} 
+ \left\{
F_{n,j}' \big[ F_{n,j}' {\mathcal F}_{n,j} + g  {\dot E}_n \big]
\right.
\\ \nonumber
&& \left.
+ F''_{n,j} {\mathcal F}^2_{n,j} -
 g\alpha \big[ P_n + {\dot E_n}\big]
\right\}
\frac{\tau_n^{3}}{6} +
\left\lbrace 
- g \alpha 
\Big( {\dot E}_n + P_n 
\Big) 
F'_{n,j} 
+ 4 F'_{n,j} F''_{n,j} {\mathcal F}^2_{n,j}
\right.
\\  \nonumber
&&
+ F_{n,j}'^{3} {\mathcal F}_{n,j} + 
g \left[
\Big(
3 F''_{n,j} {\mathcal F}_{n,j} + F_{n,j}'^2
\Big)
{\dot E}_n + \alpha^2( \dot{E}_n + 2 P_n)
\right]
+ F'''_{n,j} {\mathcal F}^3_{n,j}
\left.
\right\rbrace \frac{\tau_n^4}{24}
+O(\tau_n^5)
\qquad .
\end{eqnarray}
Eqs.~(\ref{eq:map}) and (\ref{eq:tot}) define the map we are going to
investigate in the following sections.
The time needed to reach the threshold $\tau_n$ can be determined implicitely 
from Eq.~(\ref{eq:tot}) by setting $j=1$, since by definition of the model
$u_{n, 0} \equiv 1$.

\section{Splay state solution}
\label{sec:main}

The splay state is a fixed point of the previous mapping corresponding to
a constant interspike interval $\tau = T/N$. Since the fixed point solutions
do not depend on the index $n$ they are denoted as,
\begin{equation}
\label{eq:ps}
  E_n\equiv \tilde E \ ,\quad \ P_n\equiv \tilde P \ , \quad \
  u_{n,j} \equiv  \tilde u_j\ .
\end{equation}
In order to study the dependence of the splay state on the system size $N$, it
is necessary and sufficient to formally expand the expression of the membrane
potentials as follows
\begin{equation}
\label{eq:expuN}
{\tilde u}_{j} = \sum_{h=0,4} \frac{\tilde u_{j}^{(h)}}{N^h} 
+ O\left(\frac{1}{N^{5}}\right) \, 
\end{equation}
and, analogously, for the period $T$,
\begin{equation}
\label{sviluppoT}
T=  \sum_{h=0,4}  \frac{T^{(h)}}{N^h} +O\left(\frac{1}{N^{5}}\right) \, .
\end{equation}
This expansion can be performed by exploiting the explicit dependence of $E$ and
$P$ on $N$, as detailed in Eqs.~(\ref{sviluppoE},\ref{sviluppoP}) in
Appendix \ref{one}.

Finally, by substituting the expressions
~(\ref{sviluppoT},\ref{sviluppoP},\ref{sviluppoE},\ref{sviluppoEpunto})
in Eq.~(\ref{eq:tot}) one obtains the evolution equations for the membrane
potentials 
\begin{equation}
\label{final}
\sum_{h=0,4} \frac{\tilde u_{j-1}^{(h)}-\tilde u_{j}^{(h)}}{N^h} =
\sum_{h=1,4} \frac{{\cal Q}^{(h)}}{N^h} + O\left( \frac{1}{N^5}\right) \, ,
\end{equation}
where the $\cal Q$ variables are defined in Appendix \ref{one}.

In the large $N$ limit, one can introduce the continuous spatial coordinate
$x = j/N$.  In practice, this is tantamount to write,
\begin{equation}
\label{continuo:1}
U^{(h)}(x = j/N) = {\tilde u}_{j}^{(h)}  \qquad ,\qquad h=0,\cdots,4
\qquad .
\end{equation}
It is important to stress that the event-driven neuronal evolution 
in the comoving frame implies that $U(0)=1$, i.e. the first neuron will fire at
the next step,  and $U(1)=0$, i.e. the membrane potential of the last neuron has
been just reset to zero.
This implies that $U^{(0)}(0) = 1$ and $U^{(0)}(1) = 0$, while
$U^{(h)}(0) = U^{(h)}(1) = 0$ for any $h > 0$.

Furthermore,  by expanding $U^{(h)}(x)$ around $x=j/N$, one obtains
\begin{equation}
\label{continuo:2}
\tilde u^{(h)}_{j-1} = U^{(h)}(x-1/N)= U^{(h)}(x) +
\sum_{m=1,4} \frac{1}{m\\! }\left(\frac{-1}{N}\right)^m \frac{d^m}{dx^m}
U^{(h)}(x) + O\left(\frac{1}{N^{5}}\right),
\end{equation}
By inserting this expansion into Eq.~(\ref{final}), we obtain an equation that
can be effectively split into terms of different order that will be analysed
separately.
Notice that by retaining terms of order $h$, it is possible to determine the
original variables at order $h-1$.

\subsection{Zeroth order approximation}

By assembling the first order terms, we obtain the evolution
equation for the zeroth order membrane potential, namely
\begin{equation}
\label{eq:6}
 \frac{dU^{(0)}}{dx}=-g -T^{(0)}F(U^{(0)})
\qquad .
\end{equation}
This equation is equal to the evolution equation of the membrane
potential for a constant field $E$, with $x$ playing the role of a (inverse)
time. Please notice that, up to first order, $\tilde E = 1/T^{(0)}$
(see Eq. (\ref{sviluppoE})).
An implicit and formal solution of Eq.~(\ref{eq:6}) is,
\begin{equation}
\label{eq:11}
1-x = \int_0^{U^{(0)}} \frac{d v}{g+T^{(0)}F(v)}  \, ,
\end{equation}
where we have imposed the condition $U^{(0)}(1)=0$.
However, there is a second condition
to impose, namely $U^{(0)}(0)=1$.  This second condition transforms itself
in the equation defining the interspike time interval $T^{(0)}$, when
$N \to \infty$ (i.e. in the thermodynamic limit)
\begin{equation}
\label{eq:period0}
1 = \int_0^1 \frac{d U^{(0)}}{g+T^{(0)}F(U^{(0)})}  \, .
\end{equation}
This result is, so far, quite standard and could have been easily obtained
by just assuming a constant field $E$ in equation (\ref{eq:x1}).
If we introduce the formal relation
$F'[U^{(0)}(x)]=\frac{dF(U^{(0)})}{dU^{(0)}}$ 
in  Eq. (\ref{eq:6}) we obtain
\begin{equation}
\nonumber
 \frac{dF(U^{(0)})}{g +T^{(0)}F(U^{(0)})}= -F'[U^{(0)}(x)]dx
\qquad ,
\end{equation} 
which can be easily integrated
\begin{equation}
\nonumber
\int_{F(U^{(0)}(0))}^{F(U^{(0)}(1))} \frac{dF(U^{(0)})}{g +T^{(0)}F(U^{(0)})}=
- \int_0^1 F'[U^{(0)}(x)]dx \qquad ,
\end{equation}
giving the following relation (already derived in \cite{mohanty}, by following a
different approach)
\begin{equation}
\label{rel:period}
  \frac{e^{ - T^{(0)} H(0) }}{\mathcal{F}(U(0))} =
 \frac{ e^{ - T^{(0)} H(1) }}{\mathcal{F}(U(1))} 
\quad ,
\end{equation}
where, for later convenience, we have introduced
\begin{equation}
\label{eq:defH}
H(x) = \int_0^x F'[U^{(0)}(y)] dy \quad ,
\end{equation}
and where, for the sake of simplicity, the prime denotes derivative with
respect to the variable
$U^{(0)}$ and the dependence of $F$ and $F'$ on $U^{(0)}$ has been dropped.

\subsection{First order approximation}

By collecting the terms of order $1/N^2$, one obtains
\begin{equation}
\label{eq:7}
\frac{dU^{(1)}}{dx}= -T^{(0)}F'U^{(1)} + \frac{1}{2} \frac{d^2U^{(0)}}{dx^2} -FT^{(1)} 
- \frac{1}{2}(T^{(0)})^{2}F'F - \frac{g}{2}T^{(0)}F'
\qquad .
\end{equation}

An explicit expression for the second derivative of $U^{(0)}(x)$ appearing
in Eq.~(\ref{eq:7}) can be computed by deriving Eq.~(\ref{eq:6}) with respect
to $x$.
This allows rewriting Eq.~(\ref{eq:7}) in a simplified form, namely
\begin{equation}
\label{eq:7:2}
\frac{dU^{(1)}(x)}{dx}=-U^{(1)}T^{(0)}F' - T^{(1)} F \, .
\end{equation}
By imposing $U^{(1)}(1)=0$, one obtains the general solution of
Eq.~(\ref{eq:7:2}),
\begin{equation}
\nonumber
U^{(1)}(x)= \int_x^1 du \ T^{(1)} F[U^{(0)}(u)]
 \exp \left [ T^{(0)}\Big( H(x)-H(u)\Big) \right]
\quad ,
\end{equation}
where $H(x)$ is defined by Eq.~(\ref{eq:defH}).
The further condition to be satisfied, $U^{(1)}(0)=0$,
implies $T^{(1)}=0$ and thereby we have $U^{(1)}(x)\equiv 0$,
i.e. first-order corrections vanish both for the period and the membrane
potential.

\subsection{Second order approximation}

The second order corrections can be estimated by 
assembling terms of order $1/N^3$ and by 
imposing the previously determined conditions $T^{(1)}=0$ and $U^{(1)}(x)=0$,
\begin{eqnarray}
%%\label{eq:8}
\nonumber
\hspace{-0.5cm}
\frac{dU^{(2)}}{dx} &=& -T^{(0)}F'U^{(2)}  - \frac{1}{6}\frac{d^3 U^{(0)}}{dx^3}  
-FT^{(2)}  - \frac{g^2}{6}T^{(0)}F''  
\\ \nonumber &&
- \frac{g}{6}(T^{(0)})^{2}\left[2FF'' + F'^{2}\right] 
- \frac{(T^{(0)})^{3}}{6}\left[F''F^{2} + F'^{2}F\right]
\quad .
\end{eqnarray}

Once evaluated $d^3 U^{(0)}/dx^3$ from Eq.~(\ref{eq:6}), the above ODE
reduces to
\begin{equation}
%%\label{eq:18}
\nonumber
\frac{dU^{(2)}}{dx}=-U^{(2)}T^{(0)}F' - T^{(2)}F \qquad ;
\end{equation}
which has the same structure as Eq.~(\ref{eq:7:2}). Since one has also to
impose the same boundary conditions as for the first order,
namely $U^{(2)}(0)= U^{(2)}(1)=0$,
we can conclude that $T^{(2)}=0$ and, consequently, $U^{(2)}(x) \equiv 0$. 
Therefore, second order corrections are absent too.

\subsection{Third order approximation}

By assembling terms of order $1/N^4$, once imposed
that first and second order corrections vanish, one
obtains
\begin{eqnarray}
\nonumber
&&\frac{dU^{(3)}}{dx}= -T^{(0)}F'U^{(3)} + \frac{1}{24}\frac{d^4U^{(0)}}{dx^4} -FT^{(3)} - 
\frac{g^3}{24}T^{(0)}F''' - \frac{g^2}{6}(T^{(0)})^{2}F'F''
 \\ \nonumber
&& - \frac{g^2}{8}(T^{(0)})^{2}FF''' 
-\frac{g}{24}(T^{(0)})^{3}\left[F'^{3} + 8FF'F'' + 3gF^{2}F'''\right] 
\\ \label{eq:9}
&& - \frac{(T^{(0)})^{4}}{24}FF'\left[F'^{2} 
+ 4FF'' + F^{3}F'''\right]
\quad .
\end{eqnarray}

By replacing $d^4U^{(0)}/dx^4$ with its expression derived from
Eq.~(\ref{eq:6}), equation~(\ref{eq:9})
takes the same form as in the two previous examined cases, 
namely
\begin{equation}
%%\label{eq:19}
\nonumber
\frac{dU^{(3)}}{dx}=-U^{(3)}T^{(0)}F' - T^{(3)}F \, .
\end{equation}
Therefore, we can safely conclude that third order terms vanish too.

The LIF model can be solved exactly for any value of $N$, starting from the
asymptotic value ($N\to\infty$). As shown in Appendix B, it turns out that
the leading corrections are of fourth order for both the period $T$ and the
membrane potential.

\section{Linear stability analysis}

The fixed-point analysis has revealed that the finite-size corrections to
the stationary solutions are of order $o(1/N^3)$. Since such deviations
do not affect the leading terms of the linear stability analysis
(as it can be verified {\it a posteriori}) they will be simply neglected. Therefore,
for the sake
of simplicity, from now on, $T^{(0)}$ and ${\tilde u}_{j}^{(0)}$ will be simply
referred to as $T$ and ${\tilde u}_{j}$.

The evolution rule in tangent space is obtained by differentiating
Eq.~(\ref{eq:tot}) and Eq.~(\ref{eq:map}) around the fixed point solution. 
The explicit expression of the corresponding event-driven map is reported 
in Appendix C. It consists of evolution
equations for $\delta P_n$ and $\delta E_n$ (Eqs.~(\ref{eq:delp}) and
(\ref{eq:dele})), and for $\delta u_{n,j}$ (Eq.~(\ref{eq:20})). Finally,
$\delta \tau_n$ is determined from Eq.~(\ref{deltatau}).

As usual, the eigenvalue problem can be solved by introducing the Ansatz,
\begin{equation}
\label{eq:autovalori}
\delta u_{n,j}= \mu_{k}^n\delta u_{j}  \qquad \delta P_{n}= \mu_{k}^n\delta P
\qquad  \delta E_{n}= \mu_{k}^n\delta E \qquad \delta \tau_n = \mu_k^n\delta
\tau \, ,
\end{equation}
where $\mu_k$ labels the eigenvalues, which must also be expanded as, 
\begin{equation}
\label{definizione:mu}
\mu_{k}=
{\rm e}^{i \phi_k} {\rm e}^{(\lambda_k + i \omega_k)T/N} =
{\rm e}^{i\phi_{k}}\left(1 + \sum_{h=1,3} \frac{\Gamma^{(h)}}{N^h}
 + O\left(\frac{1}{N^4} \right)
\right) \, .
\end{equation}
where $\Gamma^{(h)}$ is, in principle, a complex number and, for the sake
of simplicity, we have dropped its dependence on $k$. Finally, as already
shown, at zeroth order, the eigenvalues correspond to a pure
rotation (specified by $\phi_k$) with no expansion or contraction, i.e.
$\Gamma^{(0)}=0$.

By inserting the above Ans\"atze in the map expression
(\ref{eq:delp},\ref{eq:dele},\ref{eq:20},\ref{deltatau}), one obtains, after
eliminating $\delta P$, $\delta E$ and $\delta \tau$, a closed equation for
$\delta u_j$,
\begin{eqnarray}
\label{eq:25}
&&
{\rm e}^{i\phi_{k}}\left(1 + \frac{\Gamma^{(1)}}{N} + \frac{\Gamma^{(2)}}{N^{2}} +
\frac{\Gamma^{(3)}}{N^{3}}\right) \delta u_{j-1} =
 \left\lbrace 1 + \overline{F}'_{j}\frac{T}{N} + 
\left[\overline{F}''_{j}\overline{\mathcal{F}}_{j} + \overline{F}'^{2}_{j}\right]
\frac{T^{2}}{2N^{2}} 
\right.
\\ \nonumber
&& \left.
+ \left[\overline{F}'''_{j}\overline{\mathcal{F}}_{j}^{2} +
4\overline{F}'_{j}\overline{F}''_{j}\overline{\mathcal{F}}_{j} +
\overline{F}'^{3}_{j}\right]\frac{T^{3}}{6N^{3}}
\right\rbrace \delta u_{j} -
\left\lbrace 
\overline{\mathcal{F}}_{j} + \overline{F}'_{j}\overline{\mathcal{F}}_{j}\frac{T}{N} + 
\left[
\frac{\overline{F}''_{j}\overline{\mathcal{F}}_{j}^{2}}{2} +
\frac{\overline{F}'^{2}_{j}}{2}\overline{\mathcal{F}}_{j}
\right. \right.
\\ \nonumber
&&  \left.
+ \frac{g}{T}\alpha^{2}\frac{e^{2i\phi_{k}} +10e^{i\phi_{k}} +1}{12(e^{i\phi_{k}}-1)^{2}} 
\left( \frac{\overline{\mathcal{F}}_{j}}{\overline{\mathcal{F}}_{1}} - 1\right) 
\right]\frac{T^{2}}{N^{2}} + \left[
\frac{\overline{F}'^{3}_{j}}{6}\overline{\mathcal{F}}_{j} +
\frac{2}{3}\overline{F}'_{j}\overline{F}''_{j}\overline{\mathcal{F}}_{j}^{2} +
\frac{\overline{F}'''_{j}}{6}\overline{\mathcal{F}}_{j}^{3} 
\right.
\\ \nonumber 
&& + \left. \frac{g\alpha^{2}}{3 T^{2}} \Gamma^{(1)} \frac{2e^{2i\phi_{k}}
-3e^{i\phi_k}}{(e^{i\phi_{k}}-1)^{3}}
\left(\frac{\overline{\mathcal{F}}_{j}}{\overline{\mathcal{F}}_{1}} -1\right) +
 \frac{g}{T}\alpha^{2}\frac{e^{2i\phi_{k}} +10e^{i\phi_{k}} +1}{12(e^{i\phi_{k}}-1)^{2}} 
\frac{\overline{\mathcal{F}}_{j} }{\overline{\mathcal{F}}_{1}}(\overline{F}'_{j} -
 \overline{F}'_{1})
 \right.
\\ \nonumber
&& \left. \left.
+ \frac{g\alpha^{2}}{T}\frac{5e^{i\phi} +1}{12(e^{i\phi}-1)^{2}}
\left(\overline{F}'_{1}\frac{\overline{\mathcal{F}}_{j}}{\overline{\mathcal{F}}_{1}} -
\overline{F}'_{j}\right) +
\frac{g\alpha^{3}}{T}\frac{e^{i\phi}(e^{i\phi}+1)}{(e^{i\phi}-1)^{3}}
\left(1 - \frac{\overline{\mathcal{F}}_{j}}{\overline{\mathcal{F}}_{1}}\right)
\right] \frac{T^{3}}{N^{3}}
\right\rbrace \frac{\delta u_{1}}{\overline{\mathcal{F}}_{1}} 
\quad ,
\end{eqnarray}
that is the object of our investigation. The overline means that the function is evaluated in ${\tilde u}_j^{(0)}$,
corresponding to the infinite $N$ limit.

\subsection{Continuum limit}

Similarly to the splay-state estimation, it is convenient to take the continuum
limit. However, at variance with the previous case, now one should take in
to account also the presence of fast scales associated to the ``spatial"
dependence of $\phi_k$.

Therefore, the correct Ansatz is slightly
more complicated and we have to separate slowly and rapidly oscillating terms,
\begin{equation}\label{eq:26}
\delta u_{j} = \pi_{j} + \vartheta_{j} e^{i\phi_{k}j},
\end{equation}
where the complex exponential term accounts for the fast oscillations of the eigenvectors,
while,
\begin{equation}\nonumber
\pi_{j}= \sum_{h=0,3} \frac{\pi_{j}^{(h)}}{N^h} + O\left(\frac{1}{N^4}\right)
\qquad , \qquad
\vartheta_{j}= \sum_{h=0,3} \frac{\vartheta_{j}^{(h)}}{N^h} +
 O\left(\frac{1}{N^4}\right)
\qquad ,
\end{equation}
are slowly varying variables.

Now, we can finally introduce the continuous variable $x=j/N$, as previously
done in real space (see Eq.~(\ref{continuo:1})),
\begin{equation}
\label{eq:contt}
\Pi^{(h)}_{j}(x=\frac{j}{N})= \pi^{(h)}_{j} \qquad , \qquad
\Theta^{(h)}_{j}(x=\frac{j}{N})=\vartheta^{(h)}_{j}
\qquad ,
\end{equation}
where $h=0,\cdots,3$. This allows expanding
$\delta u_{j-1}= \pi_{j-1} +\vartheta_{j-1}e^{i\phi_{k}(j-1)}$ 
around $x=j/N$, similarly to what done in Eq.~(\ref{continuo:2}).
At variance with the computation of the fixed point, now there are also terms
like $U(1/N)$ and $\delta U(1/N)$, whose computation requires a similar
expansion but around $x=0$. By incorporating all the expansion terms within
Eq.~(\ref{eq:25}), we have finally an equation, where terms of different orders
are naturally separated from one another. The calculations are summarized
in Appendix C and the final equation is (\ref{eq:27}).
By separately treating the different orders, we obtain differential and
ordinary equations for the $\Theta$ and $\Pi$ variables.
It turns out that it is necessary to consider in parallel different orders in
the fast and slow terms to obtain $\Theta$ and $\Pi$ to the same order.
As a consequence, we will see that it is sufficient to expand $\delta U(1/N)$
up to order $O(1/N^3)$.

\subsection{Zeroth order approximation}

By assembling terms of order ${\cal O}(1/N)$  
in Eq.~(\ref{eq:27}), multiplied by the fast oscillating factor
$e^{i\phi_{k}j}$, we obtain a first-order linear differential equation for $\Theta^{(0)}$,
namely
\begin{equation}\label{eq:44}
\frac{d\Theta^{(0)}}{dx}= -\Theta^{(0)}(TF'(U(x)) - \Gamma^{(1)}) \, ,
\end{equation}
where $\Gamma^{(1)}$ is the first order correction to the Floquet exponent 
which should be determined.
It is important to remind that the prime denotes derivative with respect to the
variable $U^{(0)}$, which has been simply redifined $U$, as previously mentioned.
The solution is
\begin{equation}
\label{eq:28}
\Theta^{(0)}(x) =
 K^{(0)}\exp \left[ \Gamma^{(1)}x - T H(x)\right] \, ,
\end{equation}
where we made use of the definition (\ref{eq:defH}) and
$K^{(0)}$ is a suitable integration constant.

By assembling now the slow terms of zeroth order and reminding the
definition of $\mathcal{F}(U(x))$, we find the following algebric equation
\begin{equation}\nonumber
\Pi^{(0)}(x)(e^{i\phi} - 1)= -\left[e^{i\phi}\Theta^{(0)}(0) +
\Pi^{(0)}(0)\right]\frac{\mathcal{F}(U(x))}{\mathcal{F}(U(0))} 
\qquad .
\end{equation}
With the help of Eq.~(\ref{eq:28}), we obtain
\begin{eqnarray}
\nonumber
\Pi^{(0)}(0)&=& -\Theta^{(0)}(0) = - K^{(0)}e^{-T H(0)}  \qquad,
\\ 
%%\label{eq:29}
\nonumber
\Pi^{(0)}(x)&=& - K^{(0)}e^{-T H(0)}\frac{\mathcal{F}(U(x))}{\mathcal{F}(U(0))}
\qquad .
\end{eqnarray}
We can now impose the boundary condition
$\delta U^{(0)}(x=1)= \Theta^{(0)}(1) + \Pi^{(0)}(1)= 0$. This implies that
 \begin{equation}\nonumber
\frac{e^{-T H(1) + \Gamma^{(1)}}}{\mathcal{F}(U(1))} =
\frac{e^{-T H(0)}}{\mathcal{F}(U(0))}
\qquad .
\end{equation}
By now exploiting Eq.~(\ref{rel:period}), we find that $\Gamma^{(1)}=0$,
i.e. the Floquet exponent (both its real and its imaginary part) is equal
to zero at first order in $1/N$. Furthermore, Eq.~(\ref{eq:28}) becomes
\begin{equation}
%%\label{eq:30}
\nonumber
\Theta^{(0)}(x) = K^{(0)}e^{- TH(x)}, 
\end{equation}
i.e. the eigenvectors are independent of the phase $\phi_{k}$ and are thereby
equal to one another. In other words we are confirmed that the degeneracy
has not been removed.

\subsection{First order approximation}

By assembling the fast terms of order $1/N^{2}$ and by setting $\Gamma^{(1)}=0$, we
find that $\Theta^{(1)}$ satisfies the following first order differential equation, 
\begin{equation}\nonumber
\frac{d\Theta^{(1)}}{dx}=\Gamma^{(2)}\Theta^{(0)} - \Theta^{(1)}TF'(U(x)) 
\qquad ,
\end{equation}
whose solution is
\begin{equation}
\label{eq:31}
\Theta^{(1)}(x) = 
\left(\Gamma^{(2)}K^{(0)}x + K^{(1)}\right)e^{-T H(x)}
\qquad ,
\end{equation}
where $K^{(1)}$ is an integration constant associated with the solution of the
previous equation.

By collecting the slow terms of order $1/N$ in Eq.~(\ref{eq:27}), one
obtains the algebric equation
\begin{equation}\nonumber
\Pi^{(1)}(x)(e^{i\phi} - 1)= -\left[e^{i\phi}\Theta^{(1)}(0) +
\Pi^{(1)}(0)\right]\frac{\mathcal{F}(U(x))}{\mathcal{F}(U(0))} 
\qquad ,
\end{equation}
whose solution is,
\begin{eqnarray}
\nonumber
\Pi^{(1)}(0)&=& -\Theta^{(1)}(0) = - K^{(1)}e^{-T H(0)}
\qquad ,
\\ 
%%%\label{eq:32}
\nonumber
\Pi^{(1)}(x)&=& - K^{(1)}e^{-T H(0)}\frac{\mathcal{F}(U(x))}{\mathcal{F}(U(0))}
\qquad .
\end{eqnarray}
By imposing the boundary condition
$\delta U^{(1)}(x=1)= \Theta^{(1)}(1) + \Pi^{(1)}(1)= 0$, it is possible to
evaluate $\Gamma^{(2)}$,
\begin{equation}\nonumber
\Theta^{(1)}(1) + \Pi^{(1)}(1)= (K^{(0)}\Gamma^{(2)} + K^{(1)})e^{-T H(1)} -
 K^{(1)}e^{-T H(0)}\frac{\mathcal{F}(U(1))}{\mathcal{F}(U(0))}=0
\qquad .
\end{equation}
By again exploiting Eq.~(\ref{rel:period}), we find that $\Gamma^{(2)}=0$ and,
thereby (from Eq.~(\ref{eq:31}))
\begin{equation}
%%\label{eq:33}
\nonumber
\Theta^{(1)}(x) = K^{(1)}e^{- T H(x)} \qquad .
\end{equation}
Altogether, we can conclude that the second order correction to the Floquet
exponent vanishes as well, and one cannot remove the degeneracy among the
eigenvectors.

\subsection{Second order approximation}

By assembling fast terms of order $1/N^{3}$ appearing in Eq.~(\ref{eq:27}) and 
by setting $\Gamma^{(1)}=\Gamma^{(2)}=0$, the following first
order differential equation for $\Theta^{(2)}$ can be derived
\begin{equation}\nonumber
\frac{d\Theta^{(2)}}{dx}=\Gamma^{(3)}\Theta^{(0)} + \Theta^{(2)}TF'(U(x))
\qquad ,
\end{equation}
whose solution is
\begin{equation}
%%\label{eq:38}
\nonumber
\Theta^{(2)}(x) = 
\left(\Gamma^{(3)}K^{(0)}x + K^{(2)}\right)e^{-T H(x)} \, ,
\end{equation}
where $K^{(2)}$ is an integration constant associated with the solution of the
previous differential equation.

Furthermore, by collecting the slow terms of order $1/N^{2}$, we obtain the
algebric equation,
\begin{eqnarray}\nonumber
&& \Pi^{(2)}(x)(e^{i\phi_{k}} -1)= g\alpha^{2} T \Theta^{(0)}(0)\enskip
\frac{e^{2i\phi_k} + 10e^{i\phi_k} +1}{12(e^{i\phi_k} -1)}\enskip
\frac{F(U(0)) - F(U(x))}{[\mathcal{F}(U(0))]^{2}} 
\\ \nonumber
&& - \left[e^{i\phi_k}\Theta^{(2)}(0) + \Pi^{(2)}(0)\right]
\frac{\mathcal{F}(U(x))}{\mathcal{F}(U(0))}
\qquad .
\end{eqnarray}
By imposing that the above equation is satisfied for $x=0$, it reduces to
\begin{eqnarray}\nonumber
\Pi^{(2)}(0)&=& - \Theta^{(2)}(0) = - K^{(2)}e^{-T H(0)}
\qquad ,
\\ 
%%\label{eq:39}
\nonumber
\Pi^{(2)}(x)&=& g\alpha^{2}T \Theta^{(0)}(0)\enskip
\frac{e^{2i\phi_k} + 10e^{i\phi_k} +1}{12(e^{i\phi_k} -1)^{2}}\enskip
\frac{F(U(0)) - F(U(x))}{[\mathcal{F}(U(0))]^{2}} - \Theta^{(2)}(0)
\frac{\mathcal{F}(U(x))}{\mathcal{F}(U(0))} \, .
\end{eqnarray}

Finally, by imposing the boundary condition
$\delta U^{(2)}(x=1)= \Theta^{(2)}(1) + \Pi^{(2)}(1)= 0$,
it is possible to determine $\Gamma^{(3)}$,
\begin{equation} 
\Gamma^{(3)}=\frac{g \alpha^2}{12}T
\frac{F(U(0)) - F(U(1))}{\mathcal{F}(U(0))\mathcal{F}(U(1))}
\left(\frac{6}{1-\cos\phi_k }-1\right)
\qquad .
\label{gamma:generic}
\end{equation}
Accordingly, $\Gamma^{(3)}$ is real and depends on the difference between
$F(U(x=1))\equiv F(0)$ and $F(U(x=0)) \equiv F(1)$, confirming the numerical
findings in \cite{calamai}. Therefore, the imaginary
terms $\omega_i$ are smaller than $1/N^2$. 

In the specific example of a leaky integrate-and-fire neuron the expression
for $\Gamma^{(3)}$ reduces to 
\begin{equation}
\label{gamma:lif}
\Gamma^{(3)}=\frac{g\alpha^2}{12}T\left(2-e^{T} - e^{-T}\right)\left(
\frac{6}{1-\cos\phi_k}-1\right) \quad ,
\end{equation} 
since, by using the equations that characterize LIF neurons, the following
relation holds 
\begin{equation}\nonumber
\frac{F(U(1)) -F(U(0))}{\mathcal{F}(U(1))\mathcal{F}(U(0))} = 
\frac{1}{(a + \frac{g}{T})^{2}e^{-T}} = 
\frac{1}{(\frac{1}{1- e^{-T}})^{2}e^{-T}} = (e^{T}+ e^{-T} -2) \qquad .
\end{equation}
All in all, Eq.~(\ref{gamma:generic}) generalizes the expression 
found for the LIF model Eq.~(\ref{gamma:lif})~\cite{calamai}\footnote{In
comparing with \cite{calamai} one should pay attention to the
different normalization used here to define  $\mu_k$ in
Eq.~(\ref{definizione:mu})}.

\section{Conclusions}

We have derived analytically the short-wavelength component of
the Floquet spectrum of the splay solution in a finite, fully coupled, network
composed of {\it generic} suprathreshold pulse-coupled phase-like
neurons. This component is marginally stable in the thermodynamic limit
and thereby requires a particular care.
The analytical estimation of the long-wavelength component was previously
derived in the small-coupling limit \cite{abbott}. It would be nice to extend
such analysis to finite coupling strength, but this is a rather problematic
goal, since the eigenvalues remain finite in the thermodynamic limit and so
there is no evident smallness parameter to invoke for a safe expansion.

Our analysis has revealed that, in discontinuous velocity fields, the SW
spectrum scales as $1/N^2$, and the stability is controlled by the sign of the
difference between the velocity at reset and at threshold.
The shape of the spectrum is otherwise {\it universal}, at least for a given
choice of the post-synaptic potential. Our formalism could be easily
implemented for any pulse shape, provided that Eq.~(\ref{eq:E}) is replaced by
the appropriate evolution equation. Preliminary numerical studies anyway
suggest that different (e.g., purely exponential) pulses yield the same
scaling behaviour, but are characterized by different Floquet spectra
\cite{simona_new}.

Moreover it is worth recalling that $\delta$-like pulses in networks of LIF
neurons give rise to a different scenario, with a finite (in)stability of
the whole SW component \cite{zillmer2}. The difference is so strong that
the two scenarios cannot be reconciled even by taking the limit 
$\alpha \to \infty$ (zero pulsewidth) as the limits $N \to \infty$ and zero
pulse-width limit do not commute \cite{zillmer2}. This reveals that even the
{\it simple} construction of a general stability theory of the splay states
requires some further progress.

\begin{acknowledgements}
We thank Mathias Wolfrum for illuminating discussions
in the early stages of this study.
This research project is part of the activity
of the Joint Italian-Israeli Laboratory on Neuroscience
funded by the Italian Ministry of Foreign Affairs.
SO and AT are grateful to the Department of Physics and Astronomy
of the University of Aarhus for the hospitality during the
final write up of this manuscript and AT acknowledges
the Villum Foundation for the support received, under
the VELUX Visiting Professor Programme 2011/12, 
for his stay at the University of Aarhus.
\end{acknowledgements}
	
\appendix

\section{Fixed-point expansion (general case)}
\label{one}

A simple calculation shows that the splay state expression (\ref{eq:ps}) can
be easily obtained by first solving Eq.~\ref{eq:map} (see also \cite{zillmer2})
\be
\label{eq:eqstat}
  \tilde P=\frac{\alpha^2}{N}\frac{1}{\left(1-{\rm e}^{-\alpha T/N}
  \right)}\quad ,\quad
  \tilde E=\frac{T }{N}\frac{\tilde P}{\left( {\rm e}^{\alpha T/N}-1\right)} \ '
\ee
where $T$ is the period of the splay state, which must be determined
self-consistently.

The $1/N$ expansion of these exact expressions leads to
\begin{eqnarray}
\label{sviluppoP}
\tilde P&=&\frac{\alpha}{T^{(0)}} +
\left[\frac{\alpha}{2} - \frac{T^{(1)}}{{T^{(0)}}^2}\right] \frac{\alpha}{N} +
\left[ \frac{\alpha^2 T^{(0)}}{12} - \frac{T^{(2)}}{{T^{(0)}}^2} +
 \frac{{T^{(1)}}^2}{{T^{(0)}}^2}  \right] \frac{\alpha}{N^2} 
\\ \nonumber && 
+ \left[ \frac{\alpha^2 T^{(1)}}{12} -\frac{T^{(3)}}{{T^{(0)}}^2} +
 2\frac{T^{(1)}T^{(2)}}{{T^{(0)}}^{3}} -\frac{{T^{(1)}}^3}{{T^{(0)}}^4} 
\right] \frac{\alpha}{N^3}
 + O\left(\frac{1}{N^{4}}\right)
\qquad ,
\end{eqnarray}

\begin{eqnarray}
\label{sviluppoE}
\tilde E &=& \frac{1}{T^{(0)}} - \frac{T^{(1)}}{{T^{(0)}}^2N} + 
\left[ - \frac{\alpha^2 T^{(0)}}{12} - \frac{T^{(2)}}{{T^{(0)}}^2} +
 \frac{{T^{(1)}}^2}{{T^{(0)}}^3} \right] \frac{1}{N^2} 
\\ \nonumber && 
+ \left[ 
- \frac{\alpha^{2}}{12}T^{(1)} -\frac{T^{(3)}}{{T^{(0)}}^2} +
 2\frac{T^{(1)}T^{(2)}}{{T^{(0)}}^{3}} - \frac{{T^{(1)}}^3}{{T^{(0)}}^4}
\right] \frac{1}{N^3}+ O\left(\frac{1}{N^{4}}\right)
\qquad ,
\end{eqnarray}
\begin{equation}
\label{sviluppoEpunto}
\dot{\tilde E}= \frac{\alpha^2}{2N} +
\frac{\alpha^2 T^{(0)}}{6N^2} +  \frac{\alpha^{3}T^{(1)}}{6N^3} 
 + O\left(\frac{1}{N^{4}}\right) \, ,
\end{equation}
where we have reported also the expansion of ${\dot {\tilde E}}$ that is
necessary to pass from expression (\ref{eq:tot}) to (\ref{final}).
Please notice that while the membrane potentials and the period are expanded
up to ${\cal O}(1/N^4)$, as in (\ref{eq:expuN}) and (\ref{sviluppoT}), here we
limit the expansion to ${\cal O}(1/N^3)$ terms, since the field variables
appearing in the event-driven map are integrated over an interspike-interval
(see (\ref{eq:plus}).

To proceed further, we need also to introduce the expansions of the
velocity field and of its derivatives,
\begin{eqnarray}
%%\label{eq:5:1}
\nonumber
&&F(\tilde u_{j})= 
{\overline F}_j + {\overline F}'_i\frac{\tilde u_{j}^{(1)}}{N} +
{\overline F}'_j\frac{\tilde u_{j}^{(2)}}{N^{2}} 
+ {\overline F}'_j\frac{\tilde u_{j}^{(3)}}{N^{3}} + {\overline F}_j''
\frac{[\tilde u_{j}^{(1)}]^{2}}{2N^{2}} +
{\overline F}''_j \frac{\tilde u_{j}^{(1)}\tilde u_{j}^{(2)}}{N^{3}} 
+ {\overline F}'''_j \frac{[\tilde u_{j}^{(1)}]^{3}}{6 N^{3}} +
O\left(\frac{1}{N^{4}}\right)
\\ \nonumber
&&F'(\tilde u_{j})= {\overline F}'_j + {\overline F}''_j\frac{\tilde u_{j}^{(1)}}{N} 
+ {\overline F}''_j \frac{\tilde u_{j}^{(2)}}{N^{2}} +
{\overline F}'''_j \frac{[\tilde u_{j}^{(1)}]^{2}}{2N^{2}} + O\left(\frac{1}{N^{3}}\right) 
\\ \nonumber
&&F''(\tilde u_{j})= {\overline F}''_j + {\overline F}'''_j \frac{\tilde u_{j}^{(1)}}{N} +
O\left(\frac{1}{N^{2}}\right)
\qquad ,
\end{eqnarray}
where the overline means that the function is computed in $\tilde u_{j}^{(0)}$,
which corresponds to the infinite $N$ limit.

By replacing the membrane potentials, the period, the self-consistent fields 
and the velocity field with their expansions, the  event-driven map (\ref{eq:tot}) 
can be formally rewritten 
for the splay state as (\ref{final}) with the introduction of the following
auxiliary variables
\begin{equation}
\label{eq:q1}
{\cal Q}^{(1)} = g + T^{(0)} {\overline F}_{j}
\qquad ,
\end{equation}
\begin{equation}
\label{eq:q2}
{\cal Q}^{(2)} =
T^{(1)}{\overline F}_{j} + \left [
\tilde u^{(1)}_{j} + \frac{g}{2} +\frac{{\overline F}_{j}}{2} T^{(0)} \right ]
{\overline F}'_{j} T^{(0)}
\qquad ,
\end{equation}
\begin{eqnarray}
\nonumber
{\cal Q}^{(3)} &=& \left[
{\overline F}'_{j} \tilde u^{(2)}_{j} + \frac{{\overline F}''_{j}}{2}
 [{\tilde u}_j^{(1)} ]^{2} +
g\frac{{\overline F}''_{j}}{2} \tilde u^{(1)}_{j} +
\frac{g^{2}}{6}{\overline F}''_{j} + 
\left( \frac{2g}{3}{\overline F}''_{j}{\overline F}_{j} + {\overline F}'^2_{j}
\tilde u^{(1)}_{j} + {\overline F}''_{j} {\overline F}_{j} \tilde u^{(1)}_{j} +
\frac{g}{3} {\overline F}_{j}'^{2} \right)  \frac{T^{(0)}}{2} 
\right. 
\\ \label{e:q3}
&& \left.
+ \left( {\overline F}''_{j} {\overline F}_{j} + {\overline F}_{j}'^{2} \right)
\frac{{T^{(0)}}^{2}{\overline F}_{j}}{6} \right] T^{(0)} + \left[
\tilde u^{(1)}_{j} + \frac{g}{2} + {\overline F}_{j} T^{(0)} \right]
 T^{(1)}{\overline F}'_{j} +
T^{(2)} {\overline F}_{j}
\qquad ,
\end{eqnarray}

\begin{eqnarray}
\label{eq:q4}
&& {\cal Q}^{(4)}=  \left[ {\overline F}'_{j} \tilde u_{j}^{(3)} +
 \left( \tilde u_{j}^{(1)} + \frac{g}{2} \right) {\overline F}''_{j} \tilde u_{j}^{(2)} +
 \left( \frac{1}{6} [\tilde u_{j}^{(1)}]^{3} + \frac{g}{4} [\tilde u_{j}^{(1)}]^{2} +
 \frac{g^2}{6}  \tilde u^{(1)} + \frac{g^3}{24}
 \right) {\overline F}'''_{j} \right] T^{(0)}  
\\ \nonumber
&& + \left[
\left( {\overline F}_{j}'^2 + {\overline F}_{j} {\overline F}''_{j} \right)
\frac{\tilde u_j^{(2)}}{2} +
\left( 3{\overline F}'_{j} {\overline F}''_{j} + {\overline F}_{j} {\overline
F}'''_{j} \right) 
     \frac{[\tilde u^{(1)}_{j}]^{2}}{4} +
g \left( 2 {\overline F'}_{j} {\overline F}''_{j} + {\overline F}_{j} {\overline
F}'''_{j}\right)
    \frac{\tilde u^{(1)}_{j}}{3} +
\frac{g^2}{6} {\overline F}'_{j} {\overline F}''_{j} +
\frac{g^2}{8} {\overline F}_{j} {\overline F}'''_{j}
  \right] {T^{(0)}}^2
\\ \nonumber
&& + \left[ \big(g + 4 \tilde u^{(1)}_{j} \big) {\overline F}_{j}'^{3} +
\big(8g + 16 \tilde u^{(1)}_{j} \big) {\overline F}_{j} {\overline F}'_{j}
{\overline F}''_{j} +
  \big(3g + 4 \tilde u^{(1)}_{j} \big)  {\overline F}^{2}_{j} {\overline
  F}'''_{j}
\right] \frac{{T^{(0)}}^{3}}{24} +
\left[ {\overline F}_{i} {\overline F}_{j}'^{3} +
     4 {\overline F}^{2}_{j} {\overline F}'_{j} {\overline F}''_{j} +
     {\overline F}^{3}_{j} {\overline F}'''_{j}
\right] \frac{{T^{(0)} }^4}{24}
\\ \nonumber
&& 
+ \left[ {\overline F}'_{j} \tilde u^{(2)}_{j} + \frac{{\overline F}''_{j}}{2}
[\tilde u_j^{(1)}]^{2} +
g\frac{{\overline F}''_{j}}{2} \tilde u^{(1)}_{j} + \frac{g^{2}}{6}{\overline
F}''_{j}
\right] T^{(1)} +
\left[ \frac{2g}{3}{\overline F}''_{i}{\overline F}_{j} + {\overline F}'^2_{j}
\tilde u^{(1)}_{j} + {\overline F}''_{j} {\overline F}_{j} \tilde u^{(1)}_{j} +
\frac{g}{3} {\overline F}_{j}'^{2} \right] T^{(0)} T^{(1)} 
\\ \nonumber
&&
+ \frac{{\overline F}_{j}}{2}  \left(
{\overline F}''_{j} {\overline F}_{j} + {\overline F}_{j}'^{2} \right) 
{T^{(0)}}^{2} T^{(1)} +
\left\lbrace
\left( \tilde u^{(1)}_{j} + \frac{g}{2} \right) T^{(2)} +
\frac{{\overline F}_{j}}{2} \left[
{T^{(1)}}^2 + 2T^{(0)}T^{(2)} \right] \right\rbrace {\overline F}'_{j} +
{\overline F}_{j} T^{(3)} \, .
\end{eqnarray}

\section{Fixed-point expansion (LIF model)}
\label{appendicezero}

In the case of the LIF neuron (see Eq.~(\ref{eq:LIF})), the fixed point of the
event-driven map reads
\begin{equation}
\label{event:driven:map}
u_{i-1}= {\rm e}^{-\tau}u_i + \chi
\qquad ,
\end{equation}
where
\begin{equation}
\label{eq:45}
\chi = a(1 -e^{-\tau}) + g\frac{{\rm e}^{-\tau} 
- {\rm e}^{-\alpha\tau}}{\alpha -1}\left(E + \frac{P}{\alpha -1}\right) 
-g\frac{\tau}{\alpha -1}{\rm e}^{-\alpha\tau}P \, .
\end{equation}
Its solution is
\begin{equation}\label{eq:47}
 u_{j}= \chi \frac{1 - {\rm e}^{-NT +j\tau}}{1 - {\rm e}^{-\tau}}
\qquad .
\end{equation}
By expanding Eq.~(\ref{eq:47}) for $j=0$ and for a generic $j$, one can derive
perturbative expressions for the period $T$ and the
membrane potential, respectively.
Let us start by substituting the expressions (\ref{sviluppoT}, \ref{sviluppoP},\ref{sviluppoE}, 
\ref{sviluppoEpunto}) in Eqs.~(\ref{eq:45}). This leads to the expansion
\begin{equation}
\label{eq:chi}
\chi = \left(a + \frac{g}{T}\right)\left[\tau - \frac{\tau^{2}}{2} + \frac{\tau^{3}}{6} 
- \frac{\tau^{4}}{24}\right] + a\frac{\tau^{5}}{120} 
+ \frac{g}{T}\frac{\tau^{5}}{720}K(\alpha) + O(1/N^{4})
\qquad ,
\end{equation}
where
\begin{equation}
\label{eq:Kdef}
 K(\alpha) = \frac{360\alpha^{6} -722\alpha^{5} +363\alpha^{4} +5\alpha^{2} 
-12\alpha +6}{(\alpha -1)^{2}} 
\qquad ,
\end{equation}
accounts for the dependence on the field dynamics.
Now, with the help of Eqs.~(\ref{sviluppoT},\ref{eq:chi}) and expanding the
exponential terms up to the fourth order, we obtain a closed equation for the
interspike interval,
\begin{eqnarray} \nonumber
&&u_{0}=1=\left( a + \frac{g}{T^{(0)}} \right)\left( 1 - {\rm e}^{-T^{(0)}} \right) + 
\frac{T^{(1)}}{N}\xi(T^{(0)}) + \frac{1}{N^2}\left[ T^{(2)}\xi(T^{(0)})
+ T^{(1)}\mathcal{W}_{1}^{(2)}\right] 
\\ \nonumber &&
+ \frac{1}{N^3}\left[ T^{(3)}\xi(T^{(0)}) +
T^{(1)}\mathcal{W}_{1}^{(3)}
+ T^{(2)}\mathcal{W}_{2}^{(3)}\right] +
\frac{1}{N^4}\left[ T^{(4)}\xi(T^{(0)}) + \zeta(T^{(0)}) + T^{(1)}\mathcal{W}_{1}^{(4)} 
\right.
\\ \label{eq:48} && \left.
+ T^{(2)}\mathcal{W}_{2}^{(4)} 
+ T^{(3)}\mathcal{W}_{3}^{(4)} \right]
\qquad ,
\end{eqnarray}
where
\begin{eqnarray} 
\nonumber
&&\zeta(T^{(0)})= -(1 - e^{-T^{(0)}})\frac{g}{120}{T^{(0)}}^{3}(1 -
\frac{K(\alpha))}{6}) \quad,
\\
\nonumber
&&\xi(T^{(0)})=\left(a + \frac{g}{T^{(0)}}\right)e^{-T^{(0)}}
-\frac{g}{{T^{(0)}}^{2}} \left(1 - e^{-T^{(0)}}\right)
\quad ,
\end{eqnarray}
while $\mathcal{W}_i^{(j)}$ identifies a term of order $1/N^{j}$ that is
multiplied by $T^{(i)}$. Since, while proceeding from lower to higher-order
terms, we find that $T^{(i)}=0$ (for $i<4$), it is not necessary to give the
explicit expression of the $\mathcal{W}_i^{(j)}$ functions as they do not
contribute at all.
 
One can equivalently expand $u_{j}$
\begin{eqnarray}\nonumber
&&u_{j}= u_{j}^{(0)} + \frac{u_{j}^{(1)}}{N} + \frac{u_{j}^{(2)}}{N^{2}} + 
\frac{u_{j}^{(3)}}{N^{3}} 
+ \frac{u_{j}^{(4)}}{N^{4}} = 
\left(a + \frac{g}{T^{(0)}} \right)\left( 1 - e^{T^{(0)}(\frac{j}{N} -1)} \right) +
\frac{T^{(1)}}{N}\varsigma(T^{(0)}) 
\\ \nonumber
&& 
+ \frac{1}{N^2}\left[ T^{(2)}\varsigma(T^{(0)}) +
T^{(1)}\mathcal{Z}_{1}^{(2)}\right] + \frac{1}{N^3}\left[ T^{(3)}\varsigma(T^{(0)}) +
T^{(1)}\mathcal{Z}_{1}^{(3)} + T^{(2)}\mathcal{Z}_{2}^{(3)}\right]
\\ \label{eq:49} && 
+ \frac{1}{N^4}\left[ \varsigma(T^{(0)})T^{(4)} +
\sigma(T^{(0)}) + T^{(1)}\mathcal{Z}_{1}^{(4)}
+ T^{(2)} \mathcal{Z}_{2}^{(4)} +
T^{(3)}\mathcal{Z}_{3}^{(4)} \right]
%\\ \label{eq:48} && 
\qquad ,
\end{eqnarray}
where
\begin{eqnarray} \nonumber
&& \varsigma(T^{(0)})= -\left(a + \frac{g}{T^{(0)}}\right)e^{T^{(0)}(\frac{j}{N} -1)}
\left(\frac{j}{N} -1\right) 
- \frac{g}{{T^{(0)}}^{2}}\left[1 - e^{T^{(0)}(\frac{j}{N} -1)}\right] 
\qquad ,
\\ \nonumber
&& \sigma(T^{(0)})= \frac{g}{120}(1 
- e^{T^{(0)}(\frac{j}{N} -1)}){T^{(0)}}^{3}\left(\frac{K}{6} -1\right)
\qquad ,
\end{eqnarray}
while we do not provide explicit expressions for $\mathcal{Z}_i^{(j)}$ as
they turn out to be irrelevant.

Now we are in the position to analyse the different orders.

\subsection{Zeroth Order}
By assembling the terms of order 1 in Eq.~(\ref{eq:48}), we obtain
\begin{equation}
\nonumber
\left(a + \frac{g}{T^{(0)}}\right)(1 - e^{-T^{(0)}})=1\, .
\end{equation}
This is an implicit defintion of the asymptotic interspike time $T^{(0)}$
\begin{equation}
\nonumber
T^{(0)} =\ln \left(\frac{aT^{(0)} + g}{T^{(0)}(a-1) + g}\right) .
\end{equation}
Analogously, we can find an explicit equation for the membrane potential by
assembling the terms of order 1 in Eq.~(\ref{eq:49})
\begin{equation}
\nonumber
u_{j}^{(0)}= \left(a + \frac{g}{T^{(0)}}\right)\left[1 - e^{T^{(0)}\left(\frac{j}{N}
-1\right)}\right] 
\qquad .
\end{equation}
In the thermodinamic limit the solution for $u_{j}^{(0)}$ becomes
\begin{equation}
\nonumber
U^{(0)}(x)= \left(a + \frac{g}{T^{(0)}}\right)\left[1 - e^{T^{(0)}(x -1)}\right]
\qquad ,
\end{equation}
which coincides with Eq.~(\ref{eq:11}) with  $F=a-U^{(0)}$.

\subsection{From first to third order}
By separately assembling the terms of order $1/N^i$ (for $i=1,2,3$) in
Eq.~(\ref{eq:48}), we obtain
\begin{equation}
%%\label{eq:T1}
\nonumber
T^{(i)}\xi(T^{(0)})=0
\qquad ,
\end{equation}
which implies that $T^{(i)}=0$ since $\xi\ne 0$. Moreover, by assembling the
terms of order $1/N^i$ in Eq.~(\ref{eq:49}), we obtain
\begin{equation}
\nonumber
u_{j}^{(1)}= \varsigma(T^{(0)})T^{(i)}
\qquad ,
\end{equation}
which thereby implies that first, second and third order corrections vanish
also for the membrane potential.

\subsection{Fourth Order}
The order which reveals a different scenario is the fourth one.
By assembling the terms of order $1/N^{4}$ in Eq.~(\ref{eq:48}) we obtain
\begin{equation}
\nonumber
T^{(4)}= -\frac{\zeta(T^{(0)})}{\xi(T^{(0)})} \, ,
\end{equation}
whose explicit expression is reported in Eq.~(\ref{eq:T4LIF}).
By analogously assembling the terms of order $1/N^{4}$ in Eq.~(\ref{eq:49}), we
obtain
\begin{equation}
\nonumber
u_{j}^{(4)}= \varsigma(T^{(0)})T^{(4)} + \sigma(T^{(0)})
\qquad ,
\end{equation}
which becomes, in the thermodynamic limit,
\begin{eqnarray}
\nonumber
&&U^{(4)}(x)= -\left(a + \frac{g}{T^{(0)}}\right)T^{(4)}e^{T^{(0)}(x -1)}(x -1) 
- g\frac{T^{(4)}}{{T^{(0)}}^{2}}\left[1 - e^{T^{(0)}(x -1)}\right] 
\\ \nonumber
&&+ \frac{g}{120}(1 
- e^{T^{(0)}(x -1)}){T^{(0)}}^{3}\left(\frac{K(\alpha)}{6} -1\right)
\quad .
\end{eqnarray}

\section{Expansion in tangent space around the fixed point}

\subsection{Introduction}

The first equations of the tangent map can be determined by differentiating
Eq.~(\ref{eq:map}) and thereby expanding in powers of $\tau$ (this is equivalent
to expanding in powers of $1/N$, as the dependence of $\tau$ on $N$ would only
generate higher order terms), 
\begin{equation}
\label{eq:delp}
\delta P_{n+1}= \left(1 -\alpha\tau + \frac{\alpha^2}{2}\tau^{2} -
 \frac{\alpha^3}{6}\tau^{3} + \frac{\alpha^4}{24}\tau^{4}\right) \delta P_n  +
 \tilde P \left( -\alpha + \alpha^{2}\tau - \frac{\alpha^3}{2}\tau^{2} +
 \frac{\alpha^4}{6}\tau^{3}\right)\delta\tau_n \qquad ,
\end{equation}

\begin{eqnarray}
\label{eq:dele}
&&\delta E_{n+1}= \left(1 -\alpha\tau + \frac{\alpha^2}{2}\tau^{2} 
- \frac{\alpha^3}{6}\tau^{3} + \frac{\alpha^4}{24}\tau^{4}\right) \delta E_n +
\left(\tau -\alpha\tau^{2} + \frac{\alpha^2}{2}\tau^{3} - \frac{\alpha^3}{6}\tau^{4}\right) 
 \delta P_n 
\\ \nonumber
&& + \left[-\alpha \tilde E\left(1 -\alpha\tau + \frac{\alpha^2}{2}\tau^{2} -
\frac{\alpha^3}{6}\tau^{3}+ \frac{\alpha^4}{24}\tau^{4}\right) +
\tilde P \left(1 -2\alpha\tau + \frac{3}{2}\alpha^{2}\tau^{2} -
 \frac{2}{3}\alpha^{3}\tau^{3} + \frac{5}{24}\alpha^{4}\tau^{4} \right) \right]
 \delta\tau_n
\qquad ,
\end{eqnarray}
where the dependence of $\tau$ on $n$ has been dropped, since we are considering
a linearization around the splay state.

By further differentiating Eq.~(\ref{eq:tot}) around the fixed point solution,
one obtains
\begin{eqnarray}
\nonumber
&&\delta u_{n+1,i-1} = \delta u_{n,i} + \left(
{\overline F}'_{i}\delta u_{n,i} + g\delta E_n \right)\tau + 
\left[
{\overline F}''_{i} \overline{\mathcal{F}}_{i} \delta u_{n,i} +
{\overline F}'_{i}\left({\overline F}'_{i}\delta u_{n,i} + g\delta E_n\right) +
g \delta {\dot E}_n\right]\frac{\tau^{2}}{2} 
\\ \nonumber
&& + \left\lbrace
2g\overline{F}''_{i} \overline{\mathcal{F}}_i \delta E_n +
\left[ \overline{F}'''_{i} \overline{\mathcal{F}}_i^2 +
\overline{F}''_{i} \left(4\overline{F}'_{i}\overline{\mathcal{F}}_{i} +
 g {\dot {\tilde E}} \right)\right]\delta u_{n,i} +
 \overline{F}^2_i \left(\overline{F}'_{i}\delta u_{n,i} + g\delta E_n\right) +
 g \overline{F}'_{i} \delta {\dot E}_n 
\right.
\\ \nonumber
&& \left.
- g\alpha\left(\delta P_n + \delta {\dot E}_n\right)
\right\rbrace \frac{\tau^{3}}{6} +
\left\lbrace
\overline{\mathcal{F}}_i + \left[ \overline{F}'_{i} \overline{\mathcal{F}}_i  +
 g {\dot {\tilde E}} \right] \tau  +
\left[
\overline{F}''_{i} \overline{\mathcal{F}}_i^2 +
(\overline{F}'_{i})^{2}\overline{\mathcal{F}}_i +
g\overline{F}'_{i}{\dot {\tilde E}}
-g\alpha ({\dot {\tilde E}} + {\tilde P})
\right] \frac{\tau^{2}}{2}  
 \right.
\\ \label{eq:20}
&&\left.
+ \left[
(\overline{F}'_i)^{3} \overline{\mathcal{F}}_{i} +
4\overline{F}'_{i}\overline{F}''_{i}\overline{\mathcal{F}}^2_i+
 \overline{F}'''_{i}\overline{\mathcal{F}}_i^3 +
 g\alpha(2\alpha - \overline{F}'_{i}){\tilde P} \right]
 \frac{\tau^{3}}{6}
 \right\rbrace \delta\tau_n
\hspace{4cm}
\qquad .
\end{eqnarray}

Finally, $\delta\tau_n$ can be determined by differentiating Eq.~(\ref{eq:tot})
for $i=1$
\begin{eqnarray}
\nonumber
&& \delta\tau_{n}= -\frac{\delta E_n}{\overline{\mathcal{F}}_{1}}
\left\lbrace 
g\tau - g \left[ \frac{1}{2}(\overline{F}'_1 + \alpha) 
+ \frac{1}{\overline{\mathcal{F}}_{1}} g {\dot {\tilde E}} 
\right] \tau^2 
+ \frac{g}{\overline{\mathcal{F}}_{1}} \right.
\left[ \right.
g\alpha \Big( {\dot {\tilde E}} + \frac{{\tilde P}}{2} \Big) 
+ \frac{\overline{\mathcal{F}}_{1}}{6}
\Big( \alpha^2 - 5 \overline{F'}_1^2 - \overline{F}''_1
\overline{\mathcal{F}}_{1} + 2\alpha\overline{F}'_1 \Big) 
\\ \nonumber
&& \left. \left.+ \frac{1}{\overline{\mathcal{F}}_{1}} \Big( \overline{F}'_1
\overline{\mathcal{F}}_{1} + g {\dot {\tilde E}}  \Big)^2 \right] \tau^3
\right\rbrace  
-\frac{\delta P_n}{\overline{\mathcal{F}}_{1}}
\left\lbrace 
\frac{g}{2}\tau^2 - \frac{g}{2}\tau^3 \left[ \frac{2}{3} \Big( \alpha + \overline{F}'_1 \Big)
+ \frac{1}{\overline{\mathcal{F}}_{1}} g {\dot {\tilde E}} \right] 
\right\rbrace 
- \frac{\delta u_{n,1}}{\overline{\mathcal{F}}_{1}}
\left\lbrace 
1 - \frac{g}{\overline{\mathcal{F}}_{1}} \tau {\dot {\tilde E}} 
\right.
\\ \nonumber
&& + \tau^2 \left[ 
\frac{g\alpha}{2 \overline{\mathcal{F}}_{1}} \Big( {\dot {\tilde E}} +
{\tilde P} \Big) + \frac{1}{\overline{\mathcal{F}}_{1}^{2}}
\Big( \overline{F}'_1 \overline{\mathcal{F}}_{1} +
g {\dot {\tilde E}} \Big)^2 - \overline{F'}_1^2 - 
\frac{3}{2}\frac{g}{\overline{\mathcal{F}}_{1}} \overline{F}'_1 {\dot {\tilde E}}
\right] 
- \tau^3 \left[
\frac{1}{3\overline{\mathcal{F}}_{1}} g\alpha \Big( \alpha + \overline{F}'_1 
\Big){\tilde P}
\right.
\\ \label{deltatau}
&& \left.\left. + \frac{1}{\overline{\mathcal{F}}_{1}^{2}} g\alpha {\dot {\tilde E}}
\Big( \frac{1}{2}\overline{F}'_1 \overline{\mathcal{F}}_{1} + g {\tilde P} \Big)
+  \frac{1}{\overline{\mathcal{F}}_{1}^{3}} g^2 {\dot {\tilde E}}^3 
+ \frac{1}{\overline{\mathcal{F}}_{1}^{2}} g^2 {\dot {\tilde E}}^2 
\Big( \alpha + \overline{F}'_1 \Big) - \frac{2}{3}g \overline{F}''_1 {\dot {\tilde E}}
\right] 
\right\rbrace \qquad .
\end{eqnarray}

In order to find the Floquet eigenvalue $\mu_k$, one should
substitute the Ans\"atze (\ref{eq:autovalori}) 
into Eqs.~(\ref{eq:delp},\ref{eq:dele}). 
This allows to find explicit expressions for $\delta P$ and $\delta E$
as a function of $\mu_k$, $\tau$ and $\delta \tau$, namely
\begin{equation}
\label{eq:21}
\delta P= -\frac{\alpha^2}{\mu_{k}-1}\left[1 - 
\frac{\alpha\tau}{2} \frac{\mu_{k} +1}{\mu_{k} -1} +
\alpha^2\tau^2 M_k - \frac{\alpha^3\tau^3}{2}
\frac{\mu_{k}(\mu_{k}+1)}{(\mu_{k}-1)^{3}}\right] \frac{\delta\tau}{T}
\qquad ,
\end{equation}

\begin{equation}
\label{eq:22}
\delta E= -\frac{\alpha}{\mu_{k}-1}
\left[\frac{\alpha\tau}{2}\frac{(\mu_{k} +1)}{\mu_k-1} - 
 2\alpha^{2}\tau^{2}M_k-
 \frac{3\alpha^3\tau^3}{2}\frac{\mu_{k}(\mu_{k} +1)}{(\mu_{k} -1)^{3}}\right]
\frac{\delta\tau}{T}  \, ,
\end{equation}
where we have introduced the shorthand notation
\begin{equation}
%%\label{eq:def01}
\nonumber
M_k = \frac{\mu^{2}_{k} +10\mu_{k} +1}{12(\mu_{k} -1)^2} 
\qquad .
\end{equation}

By substituting $\delta P$ and $\delta E$, as given by (\ref{eq:21}) and
(\ref{eq:22}), into Eq.~(\ref{deltatau}), we can express $\delta\tau$ directly in
terms of $\delta u_{1}$
\begin{eqnarray}
\label{eq:23}
&&\delta\tau= 
- \left\lbrace \overline{\mathcal{F}}_{1} +
 \frac{g \alpha^2}{T} M_k \tau^2 -
\frac{g\alpha^2}{T} \left[
\frac{\overline{F}'_1}{12} \frac{(\mu_k+5)}{(\mu_{k}-1)^{2}} +
\alpha \frac{(\mu_{k}+1)}{(\mu_{k}-1)^{3}} \right] \mu_k\tau^3
\right\rbrace \frac{\delta u_{1}}{\overline{\mathcal{F}}_{1}^2} \, ,
\end{eqnarray}
where we exploited the equality
$\overline{\mathcal{F}}_{i}\equiv\overline{F}_{i} +\frac{g}{T}$ which
follows from the fact that in the thermodynamic limit $\tilde{E}=\frac{1}{T}$.

By inserting the expressions in Eqs.~(\ref{eq:21}, \ref{eq:22}, \ref{eq:23})
into Eq.~(\ref{eq:20}),
we find a single equation for the eigenvalues and eigenvectors,
\begin{eqnarray}
%%\label{eq:24}
\nonumber
&& \mu_{k}\delta u_{i-1}= \left\lbrace 1 + \overline{F}'_{i} \tau +
 \left[\overline{F}''_{i}\overline{\mathcal{F}}_{i} + 
 \overline{F}'^2_{i}\right]\frac{\tau^{2}}{2} + 
  \left[\overline{F}'''_{i}\overline{\mathcal{F}}_{i}^{2} 
+ 4 \overline{F}'_{i}\overline{F}''_{i}\overline{\mathcal{F}}_{i} + 
\overline{F}'^3_{i}\right] \frac{\tau^{3}}{6}   \right\rbrace  \delta u_{i} 
\\ \nonumber
&&  - \left\lbrace \overline{\mathcal{F}}_{i} +
\overline{F}'_{i}\overline{\mathcal{F}}_{i}\tau +
\left[
\frac{\overline{F}''_{i}}{2}\overline{\mathcal{F}}_{i}^{2} +
  \frac{\overline{F}'^2_{i}}{2}\overline{\mathcal{F}}_{i} +
\frac{g\alpha^2}{T} M_k \left( \frac{\overline{\mathcal{F}}_{i}}
                              {\overline{\mathcal{F}}_{1}} - 1 \right) 
 \right] \tau^{2} +
 \left[
 \frac{\overline{F}'^3_{i}}{6}\overline{\mathcal{F}}_{i} +
\frac{2}{3}\overline{F}'_{i}\overline{F}''_{i}\overline{\mathcal{F}}_{i}^{2} +
\frac{\overline{F}'''_{i}}{6}\overline{\mathcal{F}}_{i}^{3}  \right. \right.
\\ \nonumber
&& + \left. \left.
\frac{g\alpha^{2}}{T}\frac{5\mu_{k} +1}{12(\mu_{k}-1)^{2}} \left(
\frac{\overline{\mathcal{F}}_{i}}{\overline{\mathcal{F}}_{1}}
\overline{F}'_{1} - \overline{F}'_{i} \right) +
 \frac{g\alpha^{3}}{T}\frac{\mu_{k}(\mu_{k}+1)}{(\mu_{k}-1)^{3}} 
 \left( 1  -\frac{\overline{\mathcal{F}}_{i}}{\overline{\mathcal{F}}_{1}}
 \right) +
\frac{g\alpha^2}{T}
\frac{\overline{\mathcal{F}}_{i}}{\overline{\mathcal{F}}_{1}}
(\overline{F}'_{i} - \overline{F}'_{1}) M_k \right] \tau^3
\right\rbrace \frac{\delta u_{1}}{\overline{\mathcal{F}}_{1}} .
\end{eqnarray} 
By now substituting the $\mu_k$ expansion (\ref{definizione:mu}) and retaining
the leading terms, we obtain Eq.~(\ref{eq:25}).

\subsection{$N\to\infty$ limit}

Once the continuous variables (\ref{eq:contt}) have been introduced, it is
necessary to estimate $U(1/N)$ and $\delta U(1/N)$, by expanding such variables
around zero. By inserting the resulting expansion for $U(1/N)$ into the
expressions for
$\overline{F}_{1}$ and $\overline{\mathcal{F}}_{1}$, we obtain, respectively
\begin{eqnarray}\nonumber
&&F\left(U(\frac{1}{N})\right)
= F(U(0)) + \frac{F'(U(0))}{N}\frac{dU}{dx}\Big|_{0} + \frac{F'(U(0))}{2N^{2}}
\frac{d^{2}U}{dx^{2}}\Big|_{0} + \frac{F''(U(0))}{2}
\left(\frac{1}{N}\frac{dU}{dx}\Big|_{0}\right)^{2} 
\\ \nonumber &&
+ \frac{F''(U(0))}{2N^{3}}\frac{dU}{dx}\Big|_{0}\frac{d^{2}U}{dx^{2}}\Big|_{0} +
\frac{F'''(U(0))}{6}\left(\frac{1}{N}\frac{dU}{dx}\Big|_{0}\right)^{3} +
\frac{F'(U(0))}{6N^{3}}\frac{d^{3}U}{dx^{3}}\Big|_{0} + O(\frac{1}{N^{4}})
\qquad ,
\end{eqnarray}
\begin{eqnarray}
\nonumber 
&&\frac{1}{\mathcal{F}(U(\frac{1}{N}))}\equiv\frac{1}{F(U(\frac{1}{N})) +
\frac{g}{T}}=  \frac{1}{\mathcal{F}(U(0))} +
\frac{1}{N}\frac{TF'(U(0))}{\mathcal{F}(U(0))} +
\frac{1}{2N^{2}}\frac{[TF'(U(0))]^{2}}{\mathcal{F}(U(0))} -
\frac{1}{2N^{2}}F''(U(0))T^{2} 
\\ \nonumber &&
+ \frac{T^{3}}{6N^{3}}\left[ 
F'''(U(0))\mathcal{F}(U(0)) - 2F'(U(0))F''(U(0)) - 
\frac{g}{T^{2}}\left(\frac{F'(U(0))}{\mathcal{F}(U(0))}\right)^{2} -
5\frac{[F'(U(0))]^{3}}{\mathcal{F}(U(0))} + 6\frac{F'(U(0))}{\mathcal{F}(U(0))}
\right] + O(\frac{1}{N^{4}})\, .
\end{eqnarray}
An analogous procedure for $\delta U(1/N)$ leads to
\begin{eqnarray}
%%\label{eq:36}
\nonumber
&&\delta U(1/N)= e^{i\phi_{k}}\left[
\Theta^{(0)}(\frac{1}{N}) +
\frac{\Theta^{(1)}(\frac{1}{N})}{N} + \frac{\Theta^{(2)}(\frac{1}{N})}{N^{2}}
\right] 
+ \Pi^{(0)}(\frac{1}{N}) + \frac{\Pi^{(1)}(\frac{1}{N})}{N} 
+ \frac{\Pi^{(2)}(\frac{1}{N})}{N^{2}} + O\left(\frac{1}{N^{3}}\right)
\\ \nonumber &&
\equiv C^{(0)} + \frac{C^{(1)}}{N} + \frac{C^{(2)}}{N^{2}} + 
O\left(\frac{1}{N^{3}}\right)  \, ,
\end{eqnarray}
where $C^{(0)}$, $C^{(1)}$, $C^{(2)}$ are defined
according to the following equations
\begin{eqnarray}
%%%\label{eq:37}
\nonumber
&&C^{(0)}=e^{i\phi_{k}}\Theta^{(0)}(0) + \Pi^{(0)}(0) 
\quad ,
\\ \nonumber
&&C^{(1)}=e^{i\phi_{k}}\left[\frac{d\Theta^{(0)}}{dx}\Big|_{0} +
\Theta^{(1)}(0)\right] + \frac{d\Pi^{(0)}}{dx}\Big|_{0} +\Pi^{(1)}(0) 
\quad ,
\\ \nonumber
&&C^{(2)}=e^{i\phi_{k}}\left[\frac{1}{2}\frac{d^{2}\Theta^{(0)}}{dx^{2}}\Big|_{0} +
\frac{d\Theta^{(1)}}{dx}\Big|_{0} + \Theta^{(2)}(0)\right] +
\frac{1}{2}\frac{d^{2}\Pi^{(0)}}{dx^{2}}\Big|_{0} + \frac{d\Pi^{(1)}}{dx}\Big|_{0} +
\Pi^{(2)}(0) \, .
\end{eqnarray}
We now expand $\delta U(1/N)$ up to the order $O\left(\frac{1}{N^{3}}\right)$,
thus neglecting higher orders, because they contribute to the definition
of $\Pi$ variable and we need terms at least of order
$O\left(\frac{1}{N^{3}}\right)$ (one order lower than needed to define
$\Theta$). By inserting the Ansatz (\ref{eq:26}) and the previous
expansions in Eq.~(\ref{eq:25}), we finally obtain a closed equation
for the eigenvalues and eigenvectors,
\begin{eqnarray} \label{eq:27} 
\nonumber
&&e^{i\phi_{k}}\left\lbrace \Pi^{(0)} +
\left[\Pi^{(1)} - {\Pi^{(0)}}' + \Pi^{(0)}\Gamma^{(1)} \right] \frac{1}{N} +
\left[ \frac{{\Pi^{(0)}}''}{2} - {\Pi^{(1)}}' + \Pi^{(2)} - {\Pi^{(0)}}'\Gamma^{(1)} 
+ \Pi^{(1)}\Gamma^{(1)}
\right.\right.
\\ \nonumber
&& \left. 
 + \Pi^{(0)}\Gamma^{(2)}\right] \frac{1}{N^2} + 
 \left[\frac{{\Pi^{(1)}}''}{2} - \frac{{\Pi^{(0)}}'''}{6}  - {\Pi^{(2)}}' +
\Pi^{(3)} + \frac{{\Pi^{(0)}}''\Gamma^{(1)}}{2} 
- {\Pi^{(1)}}'\Gamma^{(1)} + \Pi^{(2)}\Gamma^{(1)}  
\right.
\\ \nonumber
&& \left.\left.
-  {\Pi^{(0)}}'\Gamma^{(2)} + \Pi^{(1)}\Gamma^{(2)} +
\Pi^{(0)}\Gamma^{(3)}\right]\frac{1}{N^3}
\right\rbrace - \Pi^{(0)} - \left[ \Pi^{(1)} + \Pi^{(0)}A^{(1)} \right]\frac{1}{N}
- \left[ \Pi^{(2)} + \Pi^{(0)}A^{(2)}
\right.
\\ \nonumber
&& \left.   
 + \Pi^{(1)}A^{(1)}\right]\frac{1}{N^2} -
\left[ \Pi^{(3)} + \Pi^{(0)}A^{(3)} 
+ \Pi^{(1)}A^{(2)} + \Pi^{(2)}A^{(1)}\right]\frac{1}{N^3}
 + C^{(0)}B^{(0)} + \left[ C^{(0)}B^{(1)} 
\right.
\\ \nonumber 
&& \left.
+ C^{(1)}B^{(0)} \right]\frac{1}{N}
 + \left[ C^{(0)}B^{(2)} + C^{(1)}B^{(1)} + C^{(2)}B^{(0)} \right]\frac{1}{N^2}
+ \frac{\mathcal{B}}{N^3} = 
e^{i\phi_{k}j} \left\lbrace 
\left[{\Theta^{(0)}}' - \Theta^{(0)}\Gamma^{(1)} 
\right. \right.
\\ \nonumber 
&& \left.
+ \Theta^{(0)}A^{(1)} \right]\frac{1}{N} +
\left[ \left(\Theta^{(0)}A^{(2)} - \frac{{\Theta^{(0)}}''}{2}\right) + {\Theta^{(1)}}' +
{\Theta^{(0)}}'\Gamma^{(1)} - \Theta^{(1)}\Gamma^{(1)} - \Theta^{(0)}\Gamma^{(2)}  
\right. 
\\ \nonumber
&& \left. 
+ \Theta^{(1)}A^{(1)}\right]\frac{1}{N^2} + 
\left[ 
\left(\frac{- {\Theta^{(1)}}''}{2} + \Theta^{(1)}A^{(2)} \right)
 + \left( \frac{{\Theta^{(0)}}'''}{6} + \Theta^{(0)}A^{(3)}\right) +
{\Theta^{(2)}}' - \frac{{\Theta^{(0)}}''\Gamma^{(1)}}{2}  
\right.
\\ && \left.\left. 
+ {\Theta^{(1)}}'\Gamma^{(1)} - \Theta^{(2)}\Gamma^{(1)}  +
{\Theta^{(0)}}'\Gamma^{(2)} -
\Theta^{(1)}\Gamma^{(2)} - \Theta^{(0)}\Gamma^{(3)}  
+ \Theta^{(2)}A^{(1)}
\right]\frac{1}{N^3} \right\rbrace \, ,  
\end{eqnarray}
where we have introduced the shorthand notation $\mathcal{B}$ in order to
characterize a term of order $O\left(\frac{1}{N^{3}}\right)$, whose
explicit expression is not necessary, since it turns out to contribute
to the definition of the $\Pi$ variable, and it is therefore one order beyond
what we need. 
Moreover, notice that the terms appearing within round brackets in the rhs of
the above equation can be shown to be zero, due to exact algebric
cancellations that emerge from the solution of the equation order by order.
Finally,
\begin{eqnarray}\nonumber
&& A^{(1)}(U(x))= TF'(U(x)), \hspace{1cm}
A^{(2)}(U(x))=\frac{T^{2}}{2}\left\lbrace F''(U(x))\mathcal{F}(U(x)) +
[F'(U(x))]^{2}\right\rbrace   
\quad ,
\\ \nonumber
&& A^{(3)}(U(x))=\frac{T^{3}}{6} \left\lbrace F'''(U(x))[\mathcal{F}(U(x))]^{2} +
4F'(U(x))F''(U(x))\mathcal{F}(U(x)) + [F'(U(x))]^{3}\right\rbrace 
\quad ,
\\ \nonumber
&& B^{(0)}(U(x))=\frac{\mathcal{F}(U(x))}{\mathcal{F}(U(0))},  \hspace{1cm}
B^{(1)}(U(x))=\left[ TF'(U(x)) + TF'(U(0)) \right]
\frac{\mathcal{F}(U(x))}{\mathcal{F}(U(0))} 
\quad ,
\end{eqnarray}

\begin{eqnarray}\nonumber
&&B^{(2)}(U(x))=T^{2}\left\lbrace -\frac{g}{T}\alpha^{2}
\enskip\frac{e^{2i\phi_{k}} +10e^{i\phi_{k}} +1}{12(e^{i\phi_{k}}-1)^{2}}\enskip
\frac{F(U(0))-F(U(x))}{[\mathcal{F}(U(0))]^{2}} +
\frac{F''(U(x))}{2}\enskip\frac{\left[\mathcal{F}(U(x))\right]^{2}}{\mathcal{F}(U(0))}
+ \right.
\\ \nonumber &&\left.
\frac{[F'(U(x))]^{2}}{2}\enskip\frac{\mathcal{F}(U(x))}{\mathcal{F}(U(0))}\right\rbrace 
+ \left\lbrace  TF'(U(x))TF'(U(0)) + \frac{[TF'(U(0))]^{2}}{2} 
\right\rbrace \frac{\mathcal{F}(U(x))}{\mathcal{F}(U(0))}  
- \frac{T^{2}}{2}F''(U(0))\mathcal{F}(U(x))
\qquad .
\end{eqnarray}

%%%%%%%%%%%%%%%%%%%%%%%%%%%%%%%%%%%%%%%%%%%%%%%%%%%%%%%%%%%%%%%%%%%%%%%%%%%%%%
%       References
%%%%%%%%%%%%%%%%%%%%%%%%%%%%%%%%%%%%%%%%%%%%%%%%%%%%%%%%%%%%%%%%%%%%%%%%%%%%%%

\end{document}